\documentstyle[preprint,aps]{revtex}
\begin{document}
\preprint{OITS-559}
\draft
\title{Amplitude Zeros in Radiative Decays of Scalar Particles}
\author{N.G. Deshpande, Xiao-Gang He and Sechul Oh}
\address{Institute of Theoretical Science\\
University of Oregon\\
Eugene, OR 97403-5203, USA}
\date{October, 1994}
\maketitle
\begin{abstract}
 We study amplitude zeros in radiative decay processes with a photon
or a gluon emission of all possible scalar particles(e.g. scalar leptoquarks)
which may interact with the usual fermions in models beyond the standard model.
For the decays with a photon emission, the amplitudes clearly exhibit
the factorization property and the differential decay rates vanish at specific
values of a certain variable which are determined only by
the electric charges of the particles involved and independent of
the particle masses and the various couplings.  For the decays with a gluon
emission,
even though the zeros are washed away, the differential decay rates still have
distinct minima.
The branching ratios as a function of leptoquark masses are presented for
the scalar leptoquark decays.  We also comment on the decays of vector
particles
into two fermions and a photon.
\end{abstract}

\newpage
\section{Introduction}

 It has been known that there is a specific angle at which the angular
distribution
of the process $q \bar q \rightarrow W \gamma$ in lowest order vanishes, so
called
the Radiation Amplitude Zero (RAZ) phenomenon, if the magnetic moment of the W
has the Standard Model (SM) value $\kappa = 1$ \cite{1}.
This phenomenon is a consequence of the factorization property of the four
particle
scattering amplitude \cite{2}.  For any gauge theory, the internal symmetry
(charge)
dependence and the polarization (spin) dependence of the scattering amplitude
can be
factorized into separate parts at tree level if one or more of the four
particles
are massless gauge bosons.  Such a characteristic phenomenon may provide a test
for
the magnetic moments of the W\cite{3} and the charge of quarks\cite{4}.  The
RAZ
has been studied in a number of processes in the standard model
and it has also been investigated for processes with charged Higgs bosons in
extensions of the standard model, such as $q \bar q \rightarrow H^-\gamma$ ,
$H^\pm \rightarrow q \bar q \gamma$ and $H^{-} \rightarrow \tilde{d}
\tilde{\bar u}
\gamma$ (where $\tilde{d} \tilde{\bar u}$ are squarks with charges $-{1 \over
3}$
and $-{2 \over 3}$ respectively) \cite{5,6,7}.

 Depending on the particle contents in the model, there are different processes
which exhibit RAZ phenomena.
The particle contents of the minimal standard $SU(3)_{C} \times SU(2)_{L}
\times
U(1)_{Y}$ model are:
(1) Gauge particles: gluon which transforms as {\bf 8} under $SU(3)_{C}$ and
does not
transform under $SU(2)_{L} \times U(1)_{Y}$.  $W^{\pm}$, $Z$ and $\gamma$ which
are
the gauge particles of $SU(2)_{L} \times U(1)_{Y}$.
(2) The Higgs particle $H$ which transforms under the SM group as (1, 2, 1).
(3) The fermions: they are left-handed quarks $Q_{L}$, right-handed up-type
quarks
$u_{R}$ and down-type quarks $d_{R}$, left-handed leptons $L_{L}$ ,and
right-handed charged leptons $e_{R}$.
Their transformation properties under the SM group are :
\begin{eqnarray}
 Q_{L}&:&(3, 2, {1 \over 3}) \ , \;\; u_{R}:(3, 1, {4 \over 3}), \;\;
 d_{R}:(3, 1, -{2 \over 3}), \nonumber \\
 L_{L}&:&(1, 2, -1), \;\;          e_{R}:(1, 1, -2).\;
\end{eqnarray}
In this model, the factorization property exists in the processes
$q \bar q \rightarrow W \gamma$ and $l \bar \nu \rightarrow W \gamma$ \cite{8}.
If the model is extended to include two or more Higgs doublets, like the
supersymmetric
SM, charged Higgs particles $H^{\pm}$ exist.
Many theories suggest the existence of new particles.
For example, in superstring-inspired $E_6$ models\cite{9}, leptoquarks which
couple to
a lepton-quark pair naturally appear as the supersymmetric partner of the
exotic
colored particle which lies in the {\bf 27} representation of $E_6$.
These particles are bosons with fractional charges and color triplets.
There are also many particles which can couple to standard fermions directly
at tree level.  With new particles there are more processes which exhibit the
RAZ
phenomenon.  It is interesting to study the RAZ in all these processes.
In Table I, we list all the scalar particles which can couple to the standard
fermions
at the tree level.

 If some of the new particles in Table I are not too heavy, they may be
produced
in collider experiments.
The scalar leptoquark production in hadron colliders or ep colliders has been
studied
in many papers\cite{10,11}.  The discovery limits depend on the strength of
the Yukawa couplings $\lambda$.  For example, it was shown that the discovery
limits
for pair production of scalar leptoquarks are $M \approx 120$ GeV at
the Fermilab Tevatron, and the limits for single production are
model-dependent\cite{10}.  In particular, an ep collider may be considered
as an ideal machine to search for leptoquarks.  At HERA (scalar) leptoquarks
with
masses as heavy as 200 GeV can be discovered even in the case of couplings as
small
as 0.01\cite{11}.  Thus, it may be possible, for instance at HERA, to detect
the RAZ
in the decays of scalar leptoquarks with relatively heavy masses if the
couplings are
not too small.

 In this paper, we study the RAZ phenomena in decay processes with a photon or
a gluon emission of all possible scalar particles that can couple to
the standard fermions at the tree level.  We study these RAZ in a
model-independent way.
We assume the existence of these particles and find the properties
related to the RAZ.  The results can be easily applied to a specific model.

 We present in Sec.II the radiative decay cases along with a photon emission
and in
Sec.III the cases along with a gluon emission.  In Sec.IV our conclusions are
summarized.

\section{THE CASES WITH THE EMISSION OF A PHOTON}

 The most general coupling of a scalar S to fermions $q_1$ and $q_2$ can be
parametrized as
\begin{eqnarray}
 {\cal L}_{int} = {\bar q_1} (A + B \gamma_5) q_2 S ,
\end{eqnarray}
where A and B are model-dependent coupling parameters.
To lowest order, there are three diagrams (Fig.1) which contribute to the
radiative decay
\begin{eqnarray}
 S(p) \rightarrow q_1(p_1) + \bar q_2(p_2) + \gamma(k).
\end{eqnarray}
The decay amplitude is given by
\begin{eqnarray}
 M = M_a + M_b + M_c \; ,
\end{eqnarray}
where $M_a, M_b$ and $M_c$ are the amplitudes corresponding to (a), (b) and (c)
of
Fig.1, respectively.  As mentioned before, there is a factorization property in
the
decay amplitude $M$ because a massless gauge particle $\gamma$ appears in the
processes.  Indeed the decay amplitude can be written in the factorized form as
\begin{eqnarray}
 M = ie ({Q_S \over p \cdot k} - {Q_1 \over p_1 \cdot k}) \bar q_1(p_1) (A + B
\gamma_5) \Pi_{\mu} q_2(p_2) \epsilon^{\mu}_{\gamma},
 \label{M1}
\end{eqnarray}
where $Q_S$ and $Q_1$ are the electric charges of $S$ and $q_1$, respectively,
and
$\epsilon^{\mu}_{\gamma}$ is the polarization vector of the photon.
$\Pi_{\mu}$ is
defined by
\begin{eqnarray}
 \Pi_{\mu} = {1 \over{2 p_2 \cdot k}} [ {p \cdot k} (2 p_{1 \mu} + \gamma_{\mu}
 {\gamma \cdot k}) - 2 p_{\mu} {p \cdot k} ].
\label{Pi1}
\end{eqnarray}
The factor $({Q_S \over p \cdot k} - {Q_1 \over p_1 \cdot k})$ causes
the RAZ phenomenon and determines the position of the zero.

 Here we would like to comment on similar decays of vector boson into two
fermions and
a photon.  The most general vector boson-fermion coupling can be parametrized
as
\begin{eqnarray}
 {\cal L}_{int} = \bar f_1 \gamma_{\mu} (a + b \gamma_5) f_2 V^{\mu},
\end{eqnarray}
where $a$ and $b$ are model-dependent coupling parameters.
For the decay $V(p) \rightarrow f_1(p_1) \bar f_2(p_2) \gamma(k)$,
there are three diagrams at tree level similar to the ones in Fig.1.
The decay amplitude can be written into the factorized form as
\begin{eqnarray}
 M = ie ({Q_V \over p \cdot k} - {Q_1 \over p_1 \cdot k}) \bar f_1(p_1)
 \Pi_{\mu \nu} (a + b \gamma_5) f_2(p_2) \epsilon^{\mu}_{\gamma}
\epsilon^{\nu}_{V}
\end{eqnarray}
where $Q_V$ and $Q_1$ are the electric charges of $V$ and $f_1$, respectively,
and
$\epsilon^{\mu}_{\gamma}$ and $\epsilon^{\nu}_{V}$ are the polarization vectors
of the photon and the vector particle, respectively.  $\Pi_{\mu \nu}$ is
defined by
\begin{eqnarray}
 \Pi_{\mu \nu} = {1 \over{2 p_2 \cdot k}} [ {p \cdot k} (2 p_{1 \mu} +
\gamma_{\mu}
 \gamma \cdot k) \gamma_{\nu}
 - 2 {p_1 \cdot k} (p_{\mu} \gamma_{\nu} + \gamma_{\mu} k_{\nu}
 - g_{\mu \nu} \gamma \cdot k) ].
\end{eqnarray}
We have used the vector-vector-photon vertex which is the same as the
$W$-$W$-photon
vertex, except for the coupling strength given by $eQ_{V}$.
The factor $({Q_V \over p \cdot k} - {Q_1 \over p_1 \cdot k})$ is the same form
as
that for the scalar decay given in eq.(\ref{M1}).  We see that the same
factorization
property exists in vector boson decays.

 Table I shows the set of possible scalar multiplets, $H_1$ through $H_{19}$,
as well as their quantum numbers under the gauge groups: color $SU(3)_C$,
$SU(2)_L$,
and hypercharge $U(1)_Y$.  We follow the labeling as in Ref.[12].  Since we do
not
include right-handed neutrino, $H_7$ and $H_{20}$ of Ref.[12] are omitted.
Also shown in the table is the fermion bilinear products
with the corresponding Yukawa couplings.

\newpage
\begin{table}
\caption{The table lists the set of scalar multiplets that can couple to
 the standard model fermions, as well as their quantum numbers
 under the gauge groups.
 The fermion bilinears to which the scalars may couple are also shown with
 the corresponding Yukawa couplings $\lambda$.
 In the table all generation indices are suppressed.}
\begin{tabular}{ccccc}
Scalars & $SU(3)$        & $SU(2)$        & $Y$
 & Fermion bilinears   \\
        & representation & representation &
 &       \\ \hline
$H_1$ & 1 & 2 & 1
 & $\lambda_{u} \bar Q_{L} u_{R}$ $\lambda_{d} \bar Q_{L} d_{R}$      \\
      &   &   &
 & $\lambda_{e} \bar L_{L} e_{R}$   \\
$H_2$ & 8 & 2 & 1
 & $\lambda_{2} \bar Q_{L} u_{R}$ $\lambda_{2}^{'} \bar Q_{L} d_{R}$  \\
$H_3$ & 3 & 2 & $7 \over 3$
 & $\lambda_{3} \bar Q_{L} e_{R}$ $\lambda_{3}^{'} \bar u_{R} L_{L}$  \\
$H_4$ & 3 & 2 & $1 \over 3$
 & $\lambda_{4}^{'} \bar d_{R} L_{L}$ \\
$H_5$ & 3 & 1 & $-{2 \over 3}$
 & $\lambda_{5}^{'} \bar Q_{L} L_{L}^{c}$       \\
      &   &   &
 & $\lambda_{5}^{''} \bar u_{R} e_{R}^{c}$  \\
$H_6$ & 3 & 3 & $-{2 \over 3}$
 & $\lambda_{6} \bar Q_{L} L_{L}^{c}$           \\
$H_8$ & 3 & 1 & $-{8 \over 3}$
 & $\lambda_{8} \bar d_{R} e_{R}^{c}$           \\
$H_9$ & 3 & 1 & $-{2 \over 3}$
 & $\lambda_{9} \bar u_{R}^{c} d_{R}$
   $\lambda_{9}^{'} \bar Q_{L}^{c} Q_{L}$       \\
$H_{10}$ & $\bar 6$ & 1 & $-{2 \over 3}$
 & $\lambda_{10} \bar u_{R}^{c} d_{R}$
   $\lambda_{10}^{'} \bar Q_{L}^{c} Q_{L}$      \\
$H_{11}$ & 3 & 3 & $-{2 \over 3}$
 & $\lambda_{11} \bar Q_{L}^{c} Q_{L}$          \\
$H_{12}$ & $\bar 6$ & 3 & $-{2 \over 3}$
 & $\lambda_{12} \bar Q_{L}^{c} Q_{L}$          \\
$H_{13}$ & 3 & 1 & $-{8 \over 3}$
 & $\lambda_{13} \bar u_{R}^{c} u_{R}$          \\
$H_{14}$ & $\bar 6$ & 1 & $-{8 \over 3}$
 & $\lambda_{14} \bar u_{R}^{c} u_{R}$          \\
$H_{15}$ & 3 & 1 & $4 \over 3$
 & $\lambda_{15} \bar d_{R}^{c} d_{R}$          \\
$H_{16}$ & $\bar 6$ & 1 & $4 \over 3$
 & $\lambda_{16} \bar d_{R}^{c} d_{R}$          \\
$H_{17}$ & 1 & 1 & $-2$
 & $\lambda_{17} \bar L_{L} L_{L}^{c}$          \\
$H_{18}$ & 1 & 3 & $-2$
 & $\lambda_{18} \bar L_{L} L_{L}^{c}$          \\
$H_{19}$ & 1 & 1 & $-4$
 & $\lambda_{19} \bar e_{R} e_{R}^{c}$
\end{tabular}
\end{table}

 As a representative example, we consider the scalar leptoquark $H_3$ in Table
I
with the Lagrangian
\begin{eqnarray}
 {\cal L}_3 &=& {{\lambda^{ij}_3} \over 2} {\bar e_i (1 - \gamma_5) u_j
H^{-5/3}_3}
 + {{\lambda^{'ij}_3} \over 2} {\bar e_i (1 + \gamma_5) u_j H^{-5/3}_3}
\nonumber \\
 &+& {{\lambda^{ij}_3} \over 2} {\bar e_i (1 - \gamma_5) d_j H^{-2/3}_3}
 + {{\lambda^{'ij}_3} \over 2} {\bar \nu_i (1 + \gamma_5) u_j H^{-2/3}_3}
 + H.C. \; ,
\end{eqnarray}
where $i$ and $j$ are generation indices, and the Yukawa couplings
$\lambda^{ij}_3$ and $\lambda^{'ij}_3$ are a priori arbitrary.

 The corresponding decay processes are:
\begin{eqnarray}
 H^{-5/3}_3(p) \rightarrow e_i(p_1) + \bar u_j(p_2) + \gamma(k) \; ,
\end{eqnarray}
\begin{eqnarray}
 H^{-2/3}_3(p) \rightarrow e_i(p_1) + \bar d_j(p_2) + \gamma(k) \; ,
\end{eqnarray}
and
\begin{eqnarray}
 H^{-2/3}_3(p) \rightarrow \nu_i(p_1) + \bar u_j(p_2) + \gamma(k) \; ,
\end{eqnarray}
where $H$, $e_i$ (or $\nu_i$) and $\bar u_j$ (or $\bar d_j$) have masses $M$,
$m_i$ and $m_j$, and electric charges $Q_H$, $Q_i$ and $Q_j$, respectively.
The lowest-order amplitude for $H^{-5/3}_3 \rightarrow e_i \bar u_j \gamma$ is:
\begin{eqnarray}
 M_3 = ie ({Q_H \over p \cdot k} - {Q_i \over p_1 \cdot k})
      \bar e_i(p_1) (A + B \gamma_5) \Pi_{\mu} u_j(p_2)
      \epsilon^{\mu}_{\gamma} ,
\label{M3}
\end{eqnarray}
where $A = ( \lambda^{ij}_3 + \lambda^{'ij}_3 ) / 2$,
$B = ( \lambda^{'ij}_3 - \lambda^{ij}_3 ) / 2$ and $\Pi_{\mu}$ defined by
eq.(\ref{Pi1}).
The amplitude can be written in a slightly different form as
\begin{eqnarray}
 M_3 = ie ({Q_i \over p_1 \cdot k} - {Q_j \over p_2 \cdot k})
      \bar e_i(p_1) (A + B \gamma_5) \Pi^{'}_{\mu} u_j(p_2)
      \epsilon^{\mu}_{\gamma} ,
\label{M2}
\end{eqnarray}
where $\Pi^{'}_{\mu}$ is defined by
\begin{eqnarray}
 \Pi^{'}_{\mu} = {1 \over 2 p \cdot k}
 [ {p_1 \cdot k} (2 p_{2 \mu} +{\gamma \cdot k} \gamma_{\mu})
 - {p_2 \cdot k} (2 p_{1 \mu} + \gamma_{\mu} {\gamma \cdot k}) ] .
\label{Pi2}
\end{eqnarray}
The amplitudes for $H^{-2/3}_3 \rightarrow e_i \bar d_j \gamma$
and $H^{-2/3}_3 \rightarrow \nu_i \bar u_j \gamma$ have the same form as the
above
one if we put $\lambda_3^{'ij}=0$ or $\lambda_3^{ij}=0$, respectively.
We see that eq.(\ref{M3}) is of the general form of eq.(\ref{M1}).
The amplitudes obviously reveal the factorization property and the factor
\( ({Q_H \over p \cdot k} - {Q_i \over p_1 \cdot k}) \) determines the position
of the
zero in each process.

 For convenience, we use the following variables introduced in Ref.[6]:
\begin{eqnarray}
 x = {2 (p_1 + p_2) \cdot k \over M^2}
\label{X}
\end{eqnarray}
and
\begin{eqnarray}
 y = {(p_1 - p_2) \cdot k \over (p_1 + p_2) \cdot k} .
\label{Y}
\end{eqnarray}
In the H rest frame, $x$ is just the scaled photon energy $2E_{\gamma}/M$.
The $x$ and $y$ limits are given by
\begin{eqnarray}
 0 \leq x \leq 1 - r
\end{eqnarray}
and for a given fixed $x$,
\begin{eqnarray}
 {{\Delta - \Lambda(x)} \over {1-x}} \leq y \leq
 {{\Delta + \Lambda(x)} \over {1-x}} ,                                \label{Y
limit}
\end{eqnarray}
where
\begin{eqnarray}
 \Lambda(x) = \sqrt{(1 - x)^2 - 2(1 - x) \epsilon + \Delta^2}.
\end{eqnarray}
The parameters $r$,$\epsilon$ and $\Delta$ above are defined as:
\begin{eqnarray}
 \mu_i &=& {m_i \over M},\;\;\; \mu_j = {m_j \over M},\;\;\;
 r = (\mu_i + \mu_j)^2,      \nonumber \\
 \epsilon &=& \mu_i^2 + \mu_j^2,\;\;\; \Delta = \mu_i^2 - \mu_j^2.
\label{Parameter}
\end{eqnarray}

 The differential decay rate for $H^{-5/3}_3 \rightarrow e_i \bar u_j \gamma$
in the H rest frame can be described as follows:
\begin{eqnarray}
{d^2 \Gamma \over {dx dy}} = {\alpha M \over 16 \pi^{2}} Q_H^2
{(y - \bar Q)^2 \over (1-y^2)}
\{ (| \lambda^{ij}_3 |^2 + | \lambda^{'ij}_3 |^2) [x + {2(1 - \epsilon) \Omega
\over x}]
 \nonumber \\
 + 4( \lambda^{ij}_3 \lambda^{'ij*}_3 + \lambda^{ij*}_3 \lambda^{'ij}_3 )
      {\mu_i \mu_j \Omega \over x} \}   \;,
\label{Gamma}
\end{eqnarray}
with $Q_H = Q_i + Q_j = -5/3$, where
\begin{eqnarray}
 \bar Q = {{Q_i - Q_j} \over {Q_i + Q_j}}
\end{eqnarray}
and
\begin{eqnarray}
 \Omega = 1 - x - 2( {\mu_i^2 \over {1 + y}} + {\mu_j^2 \over {1 - y}}) .
\end{eqnarray}
The differential decay rates for $H^{-2/3}_3 \rightarrow e_i \bar d_j \gamma$
and $H^{-2/3}_3 \rightarrow \nu_i \bar u_j \gamma$ have the same form as
eq.(\ref{Gamma}) with $Q_H = Q_i + Q_j = -2/3$, and
$\lambda_3^{'ij}=0$ (for $H^{-2/3}_3 \rightarrow e_i \bar d_j \gamma$) or
$\lambda_3^{ij}=0$ (for $H^{-2/3}_3 \rightarrow \nu_i \bar u_j \gamma$).
{}From eq.(\ref{Gamma}), it is clear that the differential decay rate vanishes
at $y = \bar Q = {{Q_i - Q_j} \over {Q_i + Q_j}}$
independent of $x$, the masses and the couplings.
For $H_3^{-5/3} \rightarrow e_i \bar u_j \gamma$, $y = \bar Q = 0.2$,
for $H_3^{-2/3} \rightarrow e_i \bar d_j \gamma$, $y = \bar Q = 2$ and
for $H_3^{-2/3} \rightarrow \nu_i \bar u_j \gamma$, $y = \bar Q = -1$.

 For a fixed $x$, the $y$ limits are given by eq.(\ref{Y limit}).
In some decay processes the corresponding values $y = \bar Q$ for the RAZ's
are not between these limit values and the RAZ's are outside the physical
region.
Due to this reason, it is impossible to detect the zeros
in some of the radiative decays.

 We summarize in Table II all possible  radiative decay processes together with
the
corresponding $\bar Q$ values.
In the table we have omitted processes of the type
$H^{0} \rightarrow q \bar q \gamma$ which does not have a RAZ with finite $\bar
Q$.
 The only detectable RAZ's occur in the radiative decays of $H_1$, $H_2$,
$H_3$, $H_6$,
$H_8$, $H_{11}$, $H_{12}$, $H_{13}$, $H_{14}$, $H_{18}$ and $H_{19}$.
In some decays, such as $H^{-} \rightarrow e \bar \nu \gamma$,
$H^{-2/3} \rightarrow \nu \bar u \gamma$ and so on, the RAZ's are just outside
the physical region at $y = \bar Q = \pm 1$.
The general form of the decays showing the detectable zeros is :
\begin{eqnarray}
 H_{\alpha}(p) \rightarrow a_{i}(p_1)
    + b_{j} (or \ \bar b_{j})(p_2) + \gamma(k) ,
\label{H}
\end{eqnarray}
where $a_{i}$ and $b_{j}$(or $\bar b_{j}$) denote (anti-)lepton or
(anti-)quark.
The indices $i$ and $j$ are generation indices
and $\alpha$ = 1,2,3,6,8,11,12,13,14,18,19.
The corresponding decay amplitude can be written as
\begin{eqnarray}
 M = ie({Q_{H} \over p \cdot k} - {Q_{i} \over p_1 \cdot k})
    \bar a_{i}(p_1) (A+B \gamma_5) \Pi_{\mu} b_{j}(p_2) \epsilon^{\mu}_{\gamma}
  \; ,
\label{M4}
\end{eqnarray}
where  $\left \{
     \begin{array}{l}
      A = B = {\lambda_{\alpha}^{ij} \over 2} \ \ $for$ \ H_1, H_2 \ $and$ \
H_6  \\
      A = -B = {\lambda_{\alpha}^{ij} \over 2} \ \ $for$ \  H_8  \\
      A = {1 \over 2}(\lambda_{\alpha}^{ij}+\lambda_{\alpha}^{ij'}) \
              $and$ \ B={1 \over
2}(\lambda_{\alpha}^{ij'}-\lambda_{\alpha}^{ij}) \ \
        $for$ \ H_3, H_{11}, H_{12}, H_{13}, H_{14}, H_{18}, H_{19}
     \end{array}
    \right. \\ $
and $\Pi_{\mu}$ is given by eq.(\ref{Pi1}).

 In Figs.2-6, we show some scalar leptoquark decays where the RAZ's are
detectable.
We normalize the differential decay rate for $H \rightarrow a_i b_j \gamma$ to
the
corresponding two body decay rate $\Gamma_0$ ($H \rightarrow a_i b_j$).
Here $\Gamma_0$ is given by
\begin{eqnarray}
 \Gamma_0
 &\equiv&  \Gamma_0(H \rightarrow a_i b_j)     \nonumber \\
 &=& {M \over 4 \pi} \sqrt{1 -2 \epsilon + \Delta^2}
 [(| \lambda^{ij} |^2 + | \lambda^{'ij} |^2) (1 - \epsilon)
 + 2( \lambda^{ij} \lambda^{'ij*} + \lambda^{ij*} \lambda^{'ij} ) \mu_i \mu_j],
\end{eqnarray}
where we have used the parameters defined by eq.(\ref{Parameter}).
In the figures we plot $(1 / \Gamma_0)(d^2 \Gamma / {dx dy})$ which shows the
position of the RAZ and also indicates the relative branching ratio of the
decay
$H \rightarrow a_i b_j \gamma$ to that of the decay $H \rightarrow a_i b_j$.
In Figs.2-4, we plot the differential decay rates versus $y$
for the scalar leptoquark decays $H^{-4/3} \rightarrow e b \gamma$ and
$H^{-5/3} \rightarrow e \bar u \gamma$.
Fig.2 shows the result for $H^{-4/3} \rightarrow e b \gamma$ with $x = 0.4$ and
$M = 200$ GeV.  The plots for $H^{-5/3} \rightarrow e \bar u \gamma$ are shown
in
Fig.3 with $\lambda_3 = \lambda^{'}_3$, $x = 0.6$ and $M = 150$ GeV, and in
Fig.4
with $\lambda_3 = 2 \lambda^{'}_3$, $x = 0.6$ and $M = 150$ GeV.  In each case
the differential decay rate vanishes at the corresponding values $y = \bar Q$
(0.5 and 0.2).

 The branching ratios versus $M$ are plotted in Figs.5 and 6.  Fig.5 is for
$ B(H^{-4/3} \rightarrow e b \gamma) =
 \Gamma(H^{-4/3} \rightarrow e b \gamma) / \Gamma_0(H^{-4/3} \rightarrow e b) $
and Fig.6 is for
$ B(H^{-5/3} \rightarrow e \bar u \gamma) =
 \Gamma(H^{-5/3} \rightarrow e \bar u \gamma) / \Gamma_0(H^{-5/3} \rightarrow e
\bar u)$
with $\lambda_3 = \lambda^{'}_3$.  In both cases, to ensure identification of
photons in the experiment, we have used the cuts on the photon energy:
$x \geq x_{cut} = {2(E_\gamma)_{cut} \over M} = 0.1$ .
We see that the ratio of the radiative decays to the two body decays are
reasonably
large($\sim$ 10$\%$ ).  Once these particles are discovered, it is possible to
study the RAZ
phenomena.

\newpage
\begin{table}
\caption{ The table shows all possible radiative decays of the scalar particles
presented in Table I.
The $\bar Q$ values for each decay are shown in parenthesis.}
\begin{tabular}{lllll}
 \hspace{1.5cm}
  $H_{1,2}^{-} \rightarrow \bar u_{i} d_{j} \gamma$ & ($\bar Q = {1 \over 3}$),
  & \hspace{2cm} $H_{1}^{-} \rightarrow e_{i} \bar \nu_{j} \gamma$
  & \hspace{0.5cm} ($\bar Q = 1$) \\
 \hspace{1.5cm}
  $H_{3}^{-5/3} \rightarrow e_{i} \bar u_{j} \gamma$    &  ($\bar Q = 0.2$),
  & \hspace{2cm} $H_{3,4}^{-2/3} \rightarrow e_{i} \bar d_{j} \gamma$
  & \hspace{0.5cm} ($\bar Q = 2$)  \\
 \hspace{1.5cm}
  $H_{3}^{-2/3} \rightarrow \nu_{i} \bar u_{j} \gamma$  & ($\bar Q = -1$),
  & \hspace{2cm} $H_{4}^{1/3} \rightarrow \nu_{i} \bar d_{j} \gamma$
  & \hspace{0.5cm} ($\bar Q = -1$) \\
 \hspace{1.5cm}
  $H_{5}^{1/3} \rightarrow \bar \nu_{i} d_{j} \gamma$   & ($\bar Q = -1$),
  & \hspace{2cm} $H_{5,6}^{-1/3} \rightarrow e_{i} u_{j} \gamma$
  & \hspace{0.5cm} ($\bar Q = 5$)    \\
 \hspace{1.5cm}
  $H_{6}^{2/3} \rightarrow \nu_{i} u_{j} \gamma$        & ($\bar Q = -1$),
  & \hspace{2cm} $H_{6}^{-1/3} \rightarrow \nu_{i} d_{j} \gamma$
  & \hspace{0.5cm} ($\bar Q = -1$)   \\
 \hspace{1.5cm}
  $H_{6,8}^{-4/3} \rightarrow e_{i} d_{j} \gamma$       & ($\bar Q = 0.5$),
  & \hspace{2cm} $H_{9,10,11,12}^{1/3} \rightarrow u_{i} d_{j} \gamma$
  & \hspace{0.5cm} ($\bar Q = 3$) \\
 \hspace{1.5cm}
  $H_{9,10,17,18}^{-} \rightarrow e_{i} \nu_{j} \gamma$ & ($\bar Q = 1$),
  & \hspace{2cm} $H_{11,12,13,14}^{4/3} \rightarrow u_{i} u_{j} \gamma$
  & \hspace{0.5cm} ($\bar Q = 0$) \\
 \hspace{1.5cm}
  $H_{11,12,15,16}^{-2/3} \rightarrow d_{i} d_{j} \gamma$  & ($\bar Q = 0$),
  & \hspace{2cm} $H_{18,19}^{-2} \rightarrow e_{i} e_{j} \gamma$
  & \hspace{0.5cm} ($\bar Q = 0$) \\
 \hspace{1.5cm} ($i$,$j$ generation indices)  &
  &  &
\end{tabular}
\end{table}

\section{THE CASES WITH THE EMISSION OF A GLUON}

 In this section we study the RAZ phenomena in scalar particle decay processes
with the emission of a gluon.  We shall find in this case that
a zero does not appear in the differential decay rate,
while the amplitude clearly shows the factorization property and reveals the
color
charge-dependent factor which is responsible for the zero.
The reason is as follows: when we calculate the differential decay rate from
the
given amplitude, we should sum over all color indices of the final particles,
because we cannot detect each color state in experiment.  The summation over
all
color indices of the product particles washes away the zero which at least
formally
appears in the amplitude.

 Now, we investigate this phenomenon specifically.  As we can see in Table I,
only
three types of decay processes may be possible: $SU(3)_{C}$ triplet-, sextet-,
and
octet-type.

 The Yukawa couplings of the scalar particles to quarks can be written as
follows.\\
$SU(3)_{C}$ triplet case:
\ ($H_3, H_4, H_5, H_6, H_7, H_8, H_9, H_{11}, H_{13}, H_{15}$)
\begin{eqnarray}
 {\cal L}_{triplet} = \lambda_t \bar q^c_i (1 + \gamma_5) q_j H_k
\epsilon^{ijk}
  + H.C. \; ,
\label{Triplet}
\end{eqnarray}
$SU(3)_{C}$ sextet case: \ ($H_{10}, H_{12}, H_{14}, H_{16}$)
\begin{eqnarray}
 {\cal L}_{sixtet} = \lambda_s \bar q^c_i (1 + \gamma_5) q^{'}_j H_{ij}
  + H.C.  \hspace{0.7cm}  (H_{ij}=-H_{ji}) \; ,
 \label{Sextet}
\end{eqnarray}
$SU(3)_{C}$ octet case: \ ($H_2$)
\begin{eqnarray}
 {\cal L}_{octet} = \lambda_o \bar q_i T^{a}_{ij} (1 + \gamma_5) q^{'}_j H^{a}
  + H.C.  \hspace{0.7cm}  (H^{a *}=H^{a}) \; ,
                          \label{Octet}
\end{eqnarray}
where $\lambda_t$, $\lambda_s$ and $\lambda_o$ denote the Yukawa couplings for
triplet, sextet and octet case, respectively.  $q$ and $q^{'}$ may denote the
different
quark flavors each other.  The indices $i$, $j$ and $k$ correspond to the quark
color
and run from 1 to 3, and $a$ corresponds to the gluon color and runs from 1 to
8.
$T^{a}_{ij}$ is the $ij$ component of $SU(3)_{C}$ color matrices $T^a$
satisfying
\begin{eqnarray}
 [ \,T^a, T^b \,] = i f_{abc} T^c \;,
\end{eqnarray}
where $f_{abc}$ are the antisymmetric $SU(3)_{C}$ structure constants.  Here
the summation
over the repeated indices are assumed.

 We consider the three-body decays of the scalar particles defined by
eqs.(\ref{Triplet}),
(\ref{Sextet}) and (\ref{Octet}) into a quark-antiquark pair (or two
antiquarks)
and a gluon.  The Feynman graphs for the triplet case are shown in Fig.7.
For the sextet and the octet cases the Feynman graphs are similar,
except that the decaying scalars have different structure of
indices such as $H_{ij}$ for the sextet and $H^{a}$ for the octet.

 The amplitudes in each case have the same form except for `the color
charge-dependent
factor' as follows:
\begin{eqnarray}
 M = ig_{s} \lambda \times
   \left( \begin{array}{c}
           color \ charge$-$ \\
           dependent \ factor
          \end{array}  \right)   \times
 \bar q^{c}_{i}(p_1) (1 + \gamma_5) \Pi_{\nu}^{'} q_j(p_2) \epsilon^{a \mu}
  \;,
\label{M5}
\end{eqnarray}
where $\lambda$ corresponds to $\lambda_{t}$ or $\lambda_{s}$ or $\lambda_{o}$
in each case and in the octet case $\bar q^{c}_{i}(p_1)$ is replaced by
$\bar q_{i}(p_1)$.  $\epsilon^{a \mu}$ is the polarization vector of gluon and
$\Pi_{\nu}^{'}$ is the one defined by eq.(\ref{Pi2}).
The color charge-dependent factors are given by
\begin{eqnarray}
  \left( \begin{array}{c}
           color \ charge$-$ \\
           dependent \ factor
          \end{array}  \right)
  = \left\{ \begin{array}{ll}
              \sum_{m} ({\epsilon_{lim} T^{a}_{mj} \over p_2 \cdot k} -
                       {\epsilon_{lmj} T^{a}_{mi} \over p_1 \cdot k})
              & $for the decay of a triplet scalar$ \ H_{l}. \\
              ({T^{a}_{nj} \delta_{mi} \over p_2 \cdot k} -
               {T^{a}_{mi} \delta_{nj} \over p_1 \cdot k})
              & $for the decay of a sextet scalar$ \ H_{mn}. \\
              \sum_{l} ({T^{b}_{il} T^{a}_{lj} \over p_2 \cdot k} +
                        {T^{a}_{il} T^{b}_{lj} \over p_1 \cdot k})
              & $for the decay of a octet scalar$ \ H^{b}.
            \end{array}  \right.
\end{eqnarray}

 As we have mentioned, the amplitudes clearly reveal the factorization property
and
the color charge-dependent factors are responsible for the RAZ.  However, to
obtain
the differential decay rates, we should sum over all color indices of the final
particles.  For example, in the case of an $SU(3)_{C}$ triplet scalar decay,
the color
charge-dependent factor which is now multipled by the complex conjugate of
itself
and summed over all color indices gives:
\begin{eqnarray}
 \sum_{ijma} |{ {\epsilon_{lim} T^{a}_{mj}} \over p_2 \cdot k } -
 { {\epsilon_{lmj} T^{a}_{mi}} \over p_1 \cdot k }|^2  \ \ (l\; \mbox{fixed})
&=& {8 \over 3} [ {1 \over (p_1 \cdot k)^2} + {1 \over (p_2 \cdot k)^2} +
    {1 \over (p_1 \cdot k) (p_2 \cdot k)} ] \nonumber \\
&=& {128 \over {3 M^4 x^2}} { {y^2 + 3} \over (1 - y^2)^2 } \;,
\label{Sum}
\end{eqnarray}
where $i$, $j$ and $m$ run from 1 to 3 and $a$ runs form 1 to 8, and we have
used
the variables $x$ and $y$ defined by eqs.(\ref{X}) and (\ref{Y}).
We note that comparing eq.(\ref{M5}) with eq.(\ref{M2}),
the differential decay rates obtained from the amplitudes of eq.(\ref{M5})
have the same form as eq.(\ref{Gamma}), except for the color charge-dependent
factors multiplied by the complex conjugates of themselves and summed over all
color
indices, such as eq.(\ref{Sum}).  However, since ${(y^2 + 3) / (1 - y^2)^2} >
0$,
it is obvious that no RAZ occurs in this case.

 The cases of an $SU(3)_{C}$ sextet and an octet scalar decay show the same
phenomena.
Explicitly, after being multiplied by the complex conjugates of themselves and
summing over all color indices, the color charge-dependent factors give:
\begin{eqnarray}
 \left\{  \begin{array}{ll}
   \sum_{ija} |{ T^{a}_{nj} \delta_{mi} \over {p_2 \cdot k} } -
     { {T^{a}_{mi} \delta_{nj}} \over {p_1 \cdot k} }|^2 \ \ (m,n\;
\mbox{fixed})
     =  {32 \over {M^4 x^2}} { 5 y^2 + 3 \over (1 - y^2)^2 } > 0  \ \ \
   & $for the sextet case$  \\
   \sum_{ijla} |{ T^{b}_{il} T^{a}_{lj} \over p_2 \cdot k } +
     { T^{a}_{il} T^{b}_{lj} \over p_1 \cdot k }|^2 \ \ (b\;\mbox{fixed})
     =  {8 \over {3 M^4 x^2}} { 9 y^2 + 7 \over (1 - y^2)^2 } > 0  \ \ \
   & $for the octet case$,
          \end{array}
 \right.
\end{eqnarray}
with $i$, $j$ and $l = 1,2,3$ and $a = 1,2, \ldots , 8$.   Fig.8 shows the
characteristic curve of the function ${(a y^2 + b) / (1 - y^2)^2} \ (a, b > 0)$
with $a = 1$ and $b = 3$.  The minimum of the function occurs at $y = 0$,
independent
of $a$ and $b$.

 Therefore, we find that in the cases of the decay along with the emission of a
gluon,
even though the RAZ's are washed away, the differential decay rates have the
characteristic property---the plot of the differential decay rates versus $y$
behaves like the curve of $(a y^2 + b) / (1 - y^2)^2$ with constants $a, b > 0$
and so has the minimum only at $y = 0$ which is independent of $x$,
the particle masses and the various couplings.

\section{CONCLUSIONS}

 We have studied the amplitude zeros in the radiative decay processes (along
with
the emission of a photon or a gluon) for all possible scalar particles which
may
interact with the standard fermions at the tree level.
For the decays with a photon emission, we have shown that the amplitudes
exhibit the expected factorization property and the differential decay rates
in the variables $x$ and $y$ vanish at $y = \bar Q$ which is independent of
$x$,
the particle masses and the various couplings $\lambda$.
The branching ratios versus various leptoquark masses were calculated and
ploted
for the scalar leptoquark decays.

 For the decays with a gluon emission, we have found that
even though the RAZ's are washed away, the differential decay rates still have
the
characteristic property that there is a minimum at $y = 0$ independent of $x$,
the particle masses and the various couplings.

\newpage
\leftline{{\Large\bf Figure captions}}

\begin{itemize}
 \item[FIG.1~.]
  {Tree level diagrams for $S \rightarrow q_1 \bar q_2 \gamma$.}

 \item[FIG.2~.]
  {$(1/ \Gamma_0)(d^{2} \Gamma / dx dy)$ versus $y$
  for $H^{-4/3} \rightarrow e b \gamma$ with $x$ = 0.4 and $M$ = 200 GeV.}

 \item[FIG.3~.]
  {$(1/ \Gamma_0)(d^{2} \Gamma / dx dy)$ versus $y$
   for $H^{-5/3} \rightarrow e \bar u \gamma$ with $x$ = 0.6 and $M$ = 150 GeV
   ($\lambda_{3} = \lambda_{3}^{'}$).}

 \item[FIG.4~.]
  {$(1/ \Gamma_0)(d^{2} \Gamma / dx dy)$ versus $y$
   for $H^{-5/3} \rightarrow e \bar u \gamma$ with $x$ = 0.6 and $M$ = 150 GeV
   ($\lambda_{3} = 2 \lambda_{3}^{'}$).}

 \item[FIG.5~.]
  {$B(H^{-4/3} \rightarrow e b \gamma)$ versus $M$ with $x \geq 0.1$.}

 \item[FIG.6~.]
  {$B(H^{-5/3} \rightarrow e \bar u \gamma)$ versus $M$ with $x \geq 0.1$ and
   $\lambda_{3} = \lambda_{3}^{'}$.}

 \item[FIG.7~.]
  {Tree level diagrams for the triplet case
   $H_{k} \rightarrow q_{i}^{c} \bar q_{j} g_{a}$ where $g_{a}$ denotes a gluon
   with color index $a$.}

 \item[FIG.8~.]
  {The plot of ${y^2 + 3 \over (1 - y^2)^2}$ versus $y$.  The minimum appears
at
   $y = 0$.}
\end{itemize}
\end{document}